\documentclass[preprint,preprintnumbers,amsmath,amssymb]{revtex4}
\usepackage{bm}

\begin{document}

\preprint{CAS-KITPC/ITP-186}

\title{A No-go Theorem Prohibiting Inflation in the Entropic Force Scenario\\ \vspace{2cm}
\vspace{1cm}}

\author{Miao Li}\email{mli@itp.ac.cn}
\author{Yi Pang}\email{yipang@itp.ac.cn}
\affiliation{Kavli Institute for Theoretical Physics, Key Laboratory
of Frontiers in Theoretical Physics, Institute of Theoretical
Physics, Chinese Academy of Sciences, Beijing 100190, People's
Republic of China\vspace{2cm}}%

\begin{abstract}

We show that to accommodate inflation in the entropic force scenario of Verlinde, it is necessary
to introduce a negative temperature on a holographic screen, this will introduce several
puzzles such as energy non-conservation. If one tries to modify the derivation of the Einstein
equations to avoid a negative temperature, we prove that it is impossible to find a proper
new definition of temperature to derive the Einstein equations.
\end{abstract}

\maketitle

Verlinde recently proposed that gravity is actually a thermodynamic
phenomenon emerging from the holographic principle \cite{verlinde},
based on an earlier observation of Jacobson \cite{jacobson} (some
other speculations on emergent gravity can be found in \cite
{Piazza:2005wm}). One of the important applications of the entropic
force scenario is to cosmology \cite{cosmology}, and in particular
to the dark energy problem \cite{liwang} if the new paradigm has
anything new to say about gravity. For other studies following
Verlinde, we refer to \cite{Smolin:2010kk,Konoplya} for an
incomplete list.

One of the authors of the present paper and Yi Wang observed in \cite{liwang} that in order to explain the current acceleration
of the universe, it is necessary to introduce a global holographic screen in addition to those
studied by Verlinde. One then natually wonders whether this is also the case for inflation.
In this letter we show that it is impossible to accommodate inflation with a single holographic
screen for inflation.

First, we present the derivation of the Einstein equations,
following Verlinde \cite{verlinde}. Choosing a local time-like
Killing vector $\xi^a$ and defining the generalized Newtonian
potential $\phi=1/2 \ln(-\xi^a\xi_a)$, the temperature on a surface
$S$ is
\begin{equation}\label{temperature}
   T={\hbar\over 2\pi}e^\phi N^a\partial_a\phi,
\end{equation}
where $N^a$ is a unit vector normal to $S$ as well as to $\xi^a$. Assuming the
equipartition theorem, the total mass on the holographic screen is
\begin{equation}\label{massofs}
   M={1\over 2}\int TdN={1\over 4\pi G}\int e^\phi \nabla\phi \cdot dS,
\end{equation}
Before proceeding we pause to note that the physics principle
requires that the temperature defined in (\ref{temperature}) must be
positive.

Utilizing the definition of Newtonian potential and applying the Stokes theorem, we arrive
at
\begin{equation}\label{ricciex}
   M={1\over 4\pi G}\int_\Sigma R_{ab}n^a\xi^b dV,
\end{equation}
where $\Sigma$ is the volume enclosed by $S$, $n^a$ is the unit future vector normal to $\Sigma$.
To derive the Eintein equations, we have to
assume that the mass measured against the Killing vector $\xi^a$ is the one given by the
so-called Tolman-Komar mass, thus
\begin{equation}\label{ricciex}
   M={1\over 4\pi G}\int_\Sigma R_{ab}n^a\xi^b dV=2\int_\Sigma \left( T_{ab}-{1\over 2}g_{ab}T\right)
   n^a\xi^b dV,
\end{equation}
Choosing an arbitrary Killing vector as well as an arbitrary $\Sigma$, we deduce the Einstein equations
$R_{ab}=8\pi G(T_{ab}-{1\over 2}g_{ab}T)$ or $R_{ab}-{1\over 2}g_{ab}R=8\pi GT_{ab}$.

Now, let us examine some details of the above derivation when it comes to inflation. For simplicity, let
us consider the case that inflation as well as quantum fluctuations against the background of inflation
can be described by the form of an ideal fluid, the stress tensor is
\begin{equation}\label{stresst}
  T_{ab}=(\rho +p)u_au_b +pg_{ab}.
\end{equation}
The Tolman-Komar stress tensor reads
\begin{equation}\label{tolmank}
   T_{ab}-{1\over 2}g_{ab}T=(\rho +p)u_au_b +{1\over 2}(\rho -p)g_{ab}.
\end{equation}
Since $\Sigma$ is arbitrary, one particular choice is the one
comoving with the fluid in which $u\cdot n=-1$ and one particular
choice of the local Killing vector also satisfies $u\cdot \xi =-1$.
We have in this case $2(T_{ab}-{1\over 2}g_{ab}T)n^a\xi^b=\rho +3p$.
We know that for the universe to be inflating this quantity is
negative, namely the Tolman-Komar mass is negative!

Thus, we are facing the problem that when the holographic screen is
comoving with the inflation fluid, the energy is negative thus
either the temperature is negative or the number of bits is
negative, the latter is unattainable so we have to assume the
temperature be negative. A negative temperature is not unfamiliar,
for instance some fifty years ago physicists had to consider this
possibility in nuclear physics (we thank Yi Wang for pointing this
to us), there, one simply takes a system of finite energy levels,
then the Boltzmann partition function is well-defined for a negative
temperature. However, when we introduce such a system into the
holographic screen, we encounter the following problems:

\noindent $\bullet$ We need to consider the holographic screen to consist of two systems, one
with a negative temperature and another with a positive temperature, since when inflation
ends radiation and matter starts to dominate.

\noindent $\bullet$ A test particle outside of the holographic screen must experience a repulsive
force, according to the formula $F=T{\Delta S\over \Delta x}$, entropy increases when the test
particle approaches the screen. Even in a system with negative temperature, entropy increases with
an increase of number of bits. Or put another way, after the test particle crosses into the screen,
we need more bits to describe the system. However, according to the equipartition theorem,
the energy of this test particle on the screen is $m={1\over 2}Tn$, $n$ is the number of bits
describing the test particle. Thus, $m$ is negative and energy is not conserved.

\noindent $\bullet$ There is also a contradiction between the
increase of entropy when the test particle crosses into the screen
and the usual wisdom that after more matter is dumped into a
horizon of an accelerating universe, the area of the horizon  actually decreases.

All of the above puzzles can be resolved at once if rather than choosing an arbitrary holographic
screen to describe inflation, we instead use a global screen as in \cite{liwang} to describe acceleration
of the Universe. However, when it comes to inflation we need to account for the local fluctuations
of the scalar fields, thus a local screen appears to be necessary. It remains to see whether
one can use a global screen to describe to the background inflation and local screens to describe
local fluctuations.

Another way to circumvent the above problem of negative temperature is to modify the derivation
of the Einstein equations. Since a negative temperature is caused by a negative Tolman-Komar
mass, we may try to replace the Tolman-Komar mass by another stress tensor
\begin{equation}\label{newstress}
   a(T_{ab}-bg_{ab}T),
\end{equation}
so long if $(1-b)\rho +3bp\ge 0$, we will not have the negative temperature problem. For
instance, for the extremal case $p=-\rho$, we need $4b\le 1$. A choice is $b={1\over 4}$ and this
choice requires the curvature part in (\ref{ricciex}) to be replaced by $R_{ab}-{1\over 4}g_{ab}R$.
Interestingly, we will not obtain the Einstein equations, rather new equations containing the traceless
part of the Einstein equations.

No matte how we modify the stress tensor, we shall start with the equipartition theorem, and use a
modified definition of temperature as well as the proposition that the number of bits propportional
to the area of the screen. Our aim is to derive an integral over the volume $\Sigma$ enclosed by screen
$S$, and the integrand must contain a term proportional to the scalar curvature (as required
by the modification of the stress tensor). In the following, we shall prove that the scalar curvature
is unattainable thus we have a proof of a no-go theorem.

To be most general, we use a rank two tensor $dS^{ab}$ to denote the area element on the surface $S$,
and the temperature vector in (\ref{temperature}) is replaced by a tensor $t_{ab}$.
We consider a generalization of equipartition relation on the
holographic screen as
\begin{equation}\label{equipartition}
    M=\frac{1}{2}\int_STdN=c\int_S t^{ab}dS_{ab},
\end{equation}
where $t^{ab}$ is antisymmetrized with respect to $ab$ and its
contraction with $dS_{ab}$ represents the temperature on the screen
times a modification of number of bits and an area element.
Components $t^{ab}$ are functions  of only the local Killing vector
$\xi$ and its higher order covariant derivatives up to order n.
Actually, the formula in (\ref{massofs}) can be recast into the form
(\ref{equipartition}), if one notices that
\begin{eqnarray}\label{}
\nonumber  \int_{S} e^\phi \nabla\phi \cdot
dS&=&\int_{S}\frac{N^{a}}{\sqrt{-\xi^c\xi_{c}}}\xi^b\nabla_a\xi_{b}dS=-\int_{S}\frac{N^{a}}{\sqrt{-\xi^c\xi_{c}}}\xi^b\nabla_b\xi_{a}dS,\\
&&=-\int_{S}N^{[a}\hat{\xi^{b]}}\nabla_b\xi_{a}dS,
\end{eqnarray}
where $\hat{\xi}^b\equiv\frac{\xi^{b}}{\sqrt{-\xi^c\xi_{c}}}$ is a time
like vector with unit norm and $[a,b]$ denotes the
antisymmetrization between $a, b$ due to the fact that
$\nabla_b\xi_{a}=\nabla_{[b}\xi_{a]}$. With the same assumption as
in \cite{verlinde} that $\hat{\xi^{b}}$ is also normal to $S$, then
$ N^{[a}\hat{\xi^{b]}}$ constitutes the bi-normal of the space like
screen $S$. Combined with the surface area $dS$,
$N^{[a}\hat{\xi^{b]}}dS$ is just dual to  $dS_{ab}$ used in
(\ref{equipartition}) and $t_{ab}$ is dual to $\nabla_a\xi_{b}$.

In the following, we will prove that when the volume of region
surrounded by $S$ goes to zero the integral (\ref{equipartition})
cannot be expressed as a volume integral
\begin{equation}\label{V}
    \int_{\Sigma}d\Sigma_a(c_1R^a_{~b}+c_2\delta^a_bR)\xi^b
\end{equation}
except when $t_{ab}$ contains a term proportional to
$\nabla_{a}\xi_{b}$, and $c_2=0$ in this case.

To begin with, we elaborate more on the property of local Killing
vector $\xi$. The local Killing vector at point p is defined by
\begin{equation}\label{}
    (\nabla_{a}\xi_{b}+\nabla_{b}\xi_{a})|_{\rm{p}}=0.
\end{equation}
It should be noted that this equation does not fix the local Killing
vector, in our proof we require further that the (n+1)-th and lower
order covariant derivatives of the local Killing vector possess the
same property as a Killing vector. Namely, the (n+1)-th and lower
order covariant derivatives of the local Killing vector can be
expressed by a linear combination of $\xi$ and $\nabla\xi$ through
$\nabla_a\nabla_b\xi_c=R_{cbad}\xi^d$. Without this requirement, the
high order covariant derivatives of the local Killing vector can
take arbitrary value, our proof becomes easier. This requirement can
be fulfilled, because one can always choose a local initial frame at
point p such that $g_{ab}|_{\rm{p}}=\eta_{ab}$ and
$\Gamma^a_{bc}|_{\rm{p}}=0$. Then at a nearby point $\rm{p}'$ , the
Killing equation can be written as
\begin{equation}\label{}
    \partial_a\xi_b+\partial_b\xi_a-\frac{2}{3}(R^c_{~abd}+R^c_{~bad})|_{\rm{p}}\delta
    x^d\xi_c+\cdots=0,
\end{equation}
where we have used the covariant expansion of the connection in the
neighborhood of p and $\delta
x^d=x^d|_{\rm{p}'}-x^d|_{\rm{p}}$. The solution of
above equation in general exists. If
we truncate the infinite expansion up to the power $(\delta x)^{\rm
n}$, then the corresponding solution of the truncated equation is
the local Killing vector we are after.

Utilizing the Stokes theorem, the right hand side of the
eq.(\ref{equipartition}) is transformed into

\begin{equation}\label{main}
   \int_S
    t^{ab}dS_{ab}=2\int_{\Sigma}\nabla_bt^{ab}d\Sigma_a,
\end{equation}
where $\Sigma$ is a three dimensional region containing point p
and enclosed by the screen. The integrand can be expressed as
\begin{equation}\label{Leib}
   \nabla_bt^{ab}=\frac{\partial
  t^{ab}}{\partial\xi_{c}}\nabla_{b}\xi_{c}+\frac{\partial
    t^{ab}}{\partial\nabla_{d}\xi_{c}}\nabla_{b}\nabla_{d}\xi_{c}+\cdots+\frac{\partial
    t^{ab}}{\partial(\nabla_{d_{\rm n}}\cdots\nabla_{d_{\rm 1}}\xi_{c})}\nabla_{b}\nabla_{d_{\rm n}}\cdots\nabla_{d_{\rm
   1}}\xi_{c}.
\end{equation}
As the volume of $\Sigma$ goes to zero, the integrand approaches
its value at point p. According to our setup, the high order
covariant derivatives of $\xi$ at point p can be replaced by a
linear combination of $\xi|_{\rm{p}}$ and $\nabla\xi|_{\rm{p}}$. We
notice that
\begin{equation}\label{}
   \nabla_{b}\nabla_{d_{\rm n}}\cdots\nabla_{d_{\rm
    1}}\xi_{c}|_{\rm{p}}=(\nabla_{b}\nabla_{d_{\rm n}}\cdots\nabla_{d_{\rm 3}}R_{cd_{\rm 1}d_{\rm
    2}a})|_{\rm{p}}\xi^{a}+\cdots,
\end{equation}
where there is a term containing the (n+1)-th order derivative of
metric which cannot be canceled by other terms on the right hand
side of eq.(\ref{Leib}), since these terms are composed by $g,
\partial g\cdots\partial^ng$. For ${\rm n}>1$, this term is linearly independent of the
Ricci tensor and the Ricci scalar containing only the second order
derivative of metric. Therefore, to generate terms containing no
$\partial^ng,n>2$, the generalized temperature $t^{ab}$ can only
depend on $\xi$ and $\nabla\xi$. Now the remaining terms are
\begin{equation}\label{T}
    \nabla_{b}t^{ab}|_{\rm{p}}=\frac{\partial
    t^{ab}}{\partial\xi_{c}}\nabla_{b}\xi_{c}+\frac{\partial
    t^{ab}}{\partial\nabla_{d}\xi_{c}}R_{cdbe}\xi^{e}|_{\rm{p}},
\end{equation}
where we have used the relation
$\nabla_{a}\nabla_{b}\xi_{c}=R_{cbad}\xi^{d}$. To understand the
above expression, we choose a frame where the local Killing vector
can be written as $\xi^{a}\partial_{a}=\partial_{t}$ with $t$
identified as the time coordinate in this new frame. Then
$\nabla_{d}\xi_{c}$ are functions of $\partial g$. We note that the
first term on the right hand
 side of eq.(\ref{T}) contains only $\partial g$. For eq.(\ref{T}) to be identified with
 $(c_1R^a_{~b}+c_2\delta^a_bR)\xi^b$, this term should vanish,
 in other words $\partial t^{ab}/\partial\xi=0$, since the relevant terms should be proportional to the
Riemann tensor.
The last term is proportional to the Riemann tensor with its
coefficients being functions of $g$ and $\partial g$. By expanding
(\ref{V}) to a polynomial composed of $g, \partial g$ and
$\partial^2g$, it is clear that each term in this expansion contains
only the second derivatives of metric. Thus $\partial
t^{ab}/\partial(\nabla_{d}\xi_{c})$ should be
    independent of $\partial g$ or $\nabla\xi$ since in (16) $\partial
t^{ab}/\partial(\nabla_{d}\xi_{c})$ is multiplied by the Riemann
tensor already containing the second derivatives of the metric. We
find that $\partial t^{ab}/\partial(\nabla_{d}\xi_{c})$ can only be
a function of metric. Inheriting $\nabla_{d}\xi_{c}$ its
antisymmetric nature about indices $d,c$, $\partial
t^{ab}/\partial(\nabla_{d}\xi_{c})$ should be antisymmetrical with
respect to $a, b$ and $d, c$. Taking into account all of these, it
is the unique possibility that $\partial
t^{ab}/\partial(\nabla_{d}\xi_{c})\propto g^{ad}g^{bc}-g^{ac}g^{bd}$
leading to the conclusion that $t_{ab}\propto\nabla_{a}\xi_{b}$.
This implies that only the Ricci tensor appears in (\ref{main}), and
$c_2$ always vanishes in (\ref{V}), this concludes our proof of the
no-go theorem.

In conclusion, we have shown that in the original derivation of the Einstein equations by Verlinde,
a negative temperature must be introduced for an accelerated expanding region, thus
introducing vexing physical problems. One may try to modify Verlinde's derivation by
modifying the definition of temperature, and we have shown that no modification is
appropriate to generate the correct Einstein equations.

\emph{Note added}: This no-go theorem is valid for
inflation models utilizing a fluid with negative Tolman-Komar mass,
it is not valid for $f(R)$ inflation. In the latter case, inflation is
driven by higher order derivative terms,
these terms may be introduced by using a modified temperature.
We thank Qing Guo Huang for a discussion on this point.

\section*{Acknowledgements}
We would like to thank Qing Guo Huang, Rong Xin Miao and Yi Wang for helpful
discussions. This work was supported by the NSFC grant
No.10535060/A050207, a NSFC group grant No.10821504 and Ministry of
Science and Technology 973 program under grant No.2007CB815401.


\begin{thebibliography}{99}

\bibitem{verlinde}E.~P.~Verlinde,
  arXiv:1001.0785 [hep-th].
  For a related work, see  T.~Padmanabhan,
  arXiv:0912.3165 [gr-qc].

\bibitem{jacobson}
T.~Jacobson,
  Phys.\ Rev.\ Lett.\  {\bf 75}, 1260 (1995)
  [arXiv:gr-qc/9504004].

\bibitem{Piazza:2005wm}
  F.~Piazza,
  Found.\ Phys.\  {\bf 40}, 239 (2010)
  [arXiv:hep-th/0506124].
  F.~Piazza,
  arXiv:0910.4677 [gr-qc].
  O.~Dreyer,
  PoS {\bf QG-PH}, 016 (2007)
  [arXiv:0710.4350 [gr-qc]].







\bibitem{cosmology}
 T.~Padmanabhan,
  arXiv:1001.3380 [gr-qc].
  F.~W.~Shu and Y.~Gong,
  arXiv:1001.3237 [gr-qc].
  R.~G.~Cai, L.~M.~Cao and N.~Ohta,
  Phys.\ Rev.\  D {\bf 81}, 061501 (2010)
  [arXiv:1001.3470 [hep-th]].


\bibitem{liwang}  M.~Li and Y.~Wang,
  Phys.\ Lett.\  B {\bf 687}, 243 (2010)
  [arXiv:1001.4466 [hep-th]].


\bibitem{Smolin:2010kk}
  L.~Smolin,
  arXiv:1001.3668 [gr-qc].
  S.~W.~Wei, Y.~X.~Liu and Y.~Q.~Wang,
  arXiv:1001.5238 [hep-th].
F.~Caravelli and L.~Modesto,
  arXiv:1001.4364 [gr-qc].
  Y.~Zhang, Y.~g.~Gong and Z.~H.~Zhu,
  arXiv:1001.4677 [hep-th].
C.~Gao,
  arXiv:1001.4585 [hep-th].
   Y.~Wang,
  arXiv:1001.4786 [hep-th].
  J.~W.~Lee, H.~C.~Kim and J.~Lee,
  arXiv:1001.5445 [hep-th].
   T.~Wang,
  arXiv:1001.4965 [hep-th].
  Y.~Ling and J.~P.~Wu,
  arXiv:1001.5324 [hep-th].
  Y.~S.~Myung,
  arXiv:1002.0871 [hep-th].
  Y.~Tian and X.~Wu,
  arXiv:1002.1275 [hep-th].
  L.~Zhao,
  arXiv:1002.0488 [hep-th].
  I.~V.~Vancea and M.~A.~Santos,
  arXiv:1002.2454 [hep-th].
   X.~G.~He and B.~Q.~Ma,
  arXiv:1003.2510 [hep-th].

\bibitem{Konoplya}
  R.~A.~Konoplya,
  arXiv:1002.2818 [hep-th].
  Y.~Zhao,
  arXiv:1002.4039 [hep-th].
   R.~Banerjee and B.~R.~Majhi,
  arXiv:1003.2312 [gr-qc].
  T.~Padmanabhan,
  arXiv:1003.5665 [gr-qc].
  S.~Ghosh,
  arXiv:1003.0285 [hep-th].







\end{thebibliography}
\end{document}